\documentclass[fleqn,usenatbib]{mnras}

\usepackage{newtxtext,newtxmath}

\usepackage[T1]{fontenc}

\DeclareRobustCommand{\VAN}[3]{#2}
\let\VANthebibliography\thebibliography
\def\thebibliography{\DeclareRobustCommand{\VAN}[3]{##3}\VANthebibliography}

\usepackage{graphicx}	
\usepackage{amsmath}	
\usepackage{mathtools}
\usepackage{pdflscape}
\usepackage{lipsum} 
\usepackage{enumitem}
\usepackage{pifont}

\usepackage{multicol}
\usepackage[dvipsnames]{xcolor}
\usepackage{comment}
\usepackage{caption}
\usepackage{subcaption}
\usepackage{rotating}

\title[Cir X-1 jet - SNR interaction]{A relativistic jet from a neutron star breaking out of its natal supernova remnant}

\author[K. V. S. Gasealahwe et al.]{K. V. S. Gasealahwe,$^{1,2}$\thanks{E-mail: kelebogile@saao.ac.za}
{K. Savard,$^3$ \thanks{E-mail: katherine.savard@physics.ox.ac.uk \newline These authors contributed equally to this work.}}
I. M. Monageng,$^{1,2}$
I. Heywood, $^3$
R. P. Fender,$^{1,3}$
P. A. Woudt,$^{1}$
\newauthor J. English, $^{4}$
J. H. Matthews, $^3$
H. Whitehead, $^3$
F. J. Cowie,$^3$
A. K. Hughes,$^{3}$
P. Saikia,$^{5}$
S. E. Motta,$^{3,6}$
\\
$^{1}$Department of Astronomy, University of Cape Town, Private Bag X3, 7701 Rondebosch, South Africa\\
$^{2}$South African Astronomical Observatory, P.O. Box 9, 7935
Observatory, South Africa \\
$^{3}$Department of Physics, University of Oxford, Denys Wilkinson Building, Keble Road, Oxford OX1 3RH, UK \\
$^{4}$ Department of Physics and Astronomy, University of Manitoba, Winnipeg, Manitoba, Canada R3T 2N2\\
$^{5}$Center for Astrophysics and Space Science (CASS), New York University Abu Dhabi, PO Box 129188, Abu Dhabi, UAE \\
$^{6}$Istituto Nazionale di Astrofisica, Osservatorio Astronomico di Brera, via E. Bianchi 46, 23807 Merate (LC), Italy \\
}

\date{Accepted XXX. Received YYY; in original form ZZZ}

\pubyear{2024}

\begin{document}
\label{firstpage}
\pagerange{\pageref{firstpage}--\pageref{lastpage}}
\maketitle

\begin{abstract}
The young neutron star X-ray binary, Cir X-1, resides within its natal supernova remnant and experiences ongoing outbursts every 16.5 days, likely due to periastron passage in an eccentric orbit. We present the deepest ever radio image of the field, which reveals relativistic jet-punched bubbles that are aligned with the mean axis of the smaller-scale jets observed close to the X-ray binary core. We are able to measure the minimum energy for the bubble, which is around $E_{min}$ = $10^{45} $ erg. The nature and morphological structure of the source were investigated through spectral index mapping and numerical simulations. The spectral index map reveals a large fraction of the nebula's radio continuum has a steep slope, associated with optically thin synchrotron emission, although there are distinct regions with flatter spectra. Our data are not sensitive enough to measure the spectral index of the protruding bubbles. We used the \textsc{pluto} code to run relativistic hydrodynamic simulations to try and qualitatively reproduce the observations with a combined supernova-plus-jet system. We are able to do so using a simplified model in which the asymmetrical bubbles are best represented by supernova explosion which is closely followed (within 100 years) by a phase of very powerful jets lasting less than 1000 years. These are the first observations revealing the initial breakout of neutron star jets from their natal supernova remnant, and further support the scenario in which Cir X-1 is a younger relation of the archetypal jet source SS433.
\end{abstract}

\begin{keywords}
radio continuum: transients -- X-rays: binaries -- stars: neutron -- stars: jets  
\end{keywords}



\section{Introduction}
Circinus X-1 (hereafter Cir X-1) is a X-ray binary (XRB) that has since its discovery in 1971 \citep{1971ApJ...169L..23M} shown conflicting characteristics. 
The discovery of X-ray bursts from the European X-ray Observatory Satellite (EXOSAT) \citep{1981SSRv...30..479T,1981SSRv...30..495D} observations taken in 1984 suggested a neutron star (NS) accretor with possible Type I bursts, however, Type II bursts could not be ruled out \citep{1986MNRAS.219..871T}. Later, \cite{2010ApJ...719L..84L} confirmed that the bursts are Type I using the observations from the Rossi X-ray Timing Explorer (RXTE) \citep{1993A&AS...97..355B} and Neil Gehrels Swift telescope (XRT) \citep[\textit{Swift};][]{2005SSRv..120..165B}, confirming the compact object is a NS. Moreover, the binary system is probably in an elliptical orbit of period 16.6\,days \citep{1976ApJ...208L..71K}.

The source appears to show mildly relativistic jets \citep{1998ApJ...506L.121F} which may be precessing close to the plane of the sky \citep{2019MNRAS.484.1672C} as well as evidence for ultrarelativistic jets with an angle close to the line of sight \citep{2004Natur.427..222F}. The X-ray behaviour of Cir X-1 
has been studied in detail during outbursts, yielding characteristics from the Hardness Intensity (HID) and Colour-Colour (CD) diagrams of both Z-sources and Atoll sources \citep{1999ApJ...517..472S,1995A&A...297..141O}. Z-source/Atoll sources are low mass -- low magnetic field NS star systems, distinguished by the characteristic patterns they trace in HIDs and CCDs during outbursts \citep{1989A&A...225...79H}. \cite{1994ApJS...92..511V} predicted that Cir X-1 may be an Atoll that shows Z-source behaviour during outbursts but \cite{1999ApJ...517..472S} confirmed complete Z-track behaviour. Therefore, Cir X-1 exhibits properties consistent with both Z and Atoll sources, indicating it is likely a low-mass X-ray binary. And, although the Type I bursts observed from Cir X-1 are consistent with accretion from an H/He enriched companion, the age and nature of the binary system has been a puzzling factor due to the association with the extended nebulae in the region of source.

\cite{1975Natur.254..674C} suggested that the X-ray binary was born in the nearby supernova remnant G321.9-0.3 (see Fig.~\ref{fig:roughIntensity}.) moving away from it at high speed due to a large kick in the natal supernova explosion. Later, distance and age estimates for both were shown to be broadly consistent. \cite{1986Natur.324..233H} first reported the smaller diffuse radio nebula centred on the X-ray binary itself and speculated that the southern extension of this nebula towards G 321.9-0.3 supported the hypothesis of \cite{1975Natur.254..674C}. \cite{1993MNRAS.261..593S} reported radio imaging from the Australia Telescope Compact Array (ATCA) of the Cir X-1 nebula and argued that apparently swept-back radio jets further supported the \cite{1975Natur.254..674C} hypothesis. During the 1990s further detailed radio observations revealed a smaller jet-like structures within the nebula (e.g. \citealt{1998ApJ...506L.121F}), and the association with G321.9-0.3 was widely accepted.
\\

However, \cite{2002A&A...386..487M} reported strong upper limits on the proper motion of the optical counterpart using Hubble Space Telescope (HST) observations separated by over 8 years, ruling out an association between G321.9-0.3 and the X-ray binary. Focus returned to understanding the nature of the Cir X-1 nebula itself and associated jets, with increasingly detailed radio images presented variously in \cite{2008MNRAS.390..447T} (which also presented proper motion limits on the core consistent with the HST results of \citealt{2002A&A...386..487M}), \cite{2012MNRAS.419..436C} and \cite{2019MNRAS.484.1672C}. \cite{2019MNRAS.484.1672C} reported the first detailed study into whether or not the jets were precessing, as hinted at in previous more sparse comparisons of jet images. In the meantime, our understanding of the relationship between the X-ray binary and the nebula had also changed. \cite{2013ApJ...779..171H} used {\em Chandra} observations to confirm that the Cir X-1 nebula is the natal supernova remnant of Cir X-1. The authors determined a maximum age of 4600\,years for the supernova remnant, thus making Cir X-1 the youngest known XRB. The high-resolution X-ray imaging available with {\em Chandra} also revealed X-ray structures, within the outer shell of the nebula and aligned with the radio jets \citep{2009MNRAS.397L...1S,2010ApJ...719L.194S}. Therefore, prior to the new MeerKAT imaging and analysis presented here, the 
consensus was that Cir X-1 was a very young neutron star X-ray binary. \cite{2016MNRAS.456..347J} propose that the companion is an unusual "puffed up" star as a result of heating and the deposition of energy from the supernova explosion. Furthermore, \cite{2013ApJ...779..171H} argue a likely high mass companion following a core collapse supernova explosion. However, the Z-source/Atoll nature of the source supports the alternative evolved low-mass companion scenario the authors discuss.  \cite{2015ApJ...806..265H} determined the position to be at a distance of $9.4^{+0.8}_{-1.0}$\,kpc. 

In this paper, we present new deep imaging of Cir X-1 with the MeerKAT radio telescope. Due to the striking resemblance, we named the nebula the `Africa nebula'. The imaging results are supported by detailed numerical simulations, which demonstrate our knowledge of the source was yet an incomplete picture. 

\section{MeerKAT observations}
MeerKAT hosted a large monitoring campaign for Cir X-1, observing the source daily for 34 epochs at a central frequency of 1.28\,GHz. This was done through the ThunderKAT Large Survey Programme \citep{2016mks..confE..13F}. Although our MeerKAT data were collected to monitor outbursts over time, we do not present the time-series analysis and focus instead on the deep broadband radio image of Cir X-1 and the surrounding field. A total of 8.5\,hrs was spent observing, using an average of 61 dishes for each observation. J1939-6342 was the primary calibrator (flux, bandpass) and the secondary (phase) calibrator was J1424-4913. The semi-automated pipeline, {\textsc{oxkat}}\footnote{for more details see, \url{https://github.com/IanHeywood/oxkat}} \citep{2020ascl.soft09003H} was used to reduce the data. Each epoch was imaged using a modified \textsc{oxkat} recipe for this field because of the multiple bright large-scale structures within the field. The 1GC, Flag-only, 2GC-multiscale and 3GC peeling scripts were used in succession to image each observation (see \citealt{2022MNRAS.509.2150H} for a detailed description of the calibration and imaging procedures). In order to improve the image, two of the bright background sources were peeled ($>$3\,Jy in the North and South region of the field). To create the deep image we made a clean mask with a pixel threshold of 0.002, summing the 30 epochs (total of 7.5\,hrs) of the best images (such that the epochs 4, 11, 14, 15, were excluded due to bad artefacts). We then ran a \textsc{wsclean} job with the new mask on the 30 measurements sets, with multiscale parameters 0,3,9,18,36, weighted Briggs of -0.3 and a circular beam to create the final image.

\section{Results and Analyses}\label{sec:Results}
\begin{figure*}
    \centering
    \includegraphics[scale=0.3]{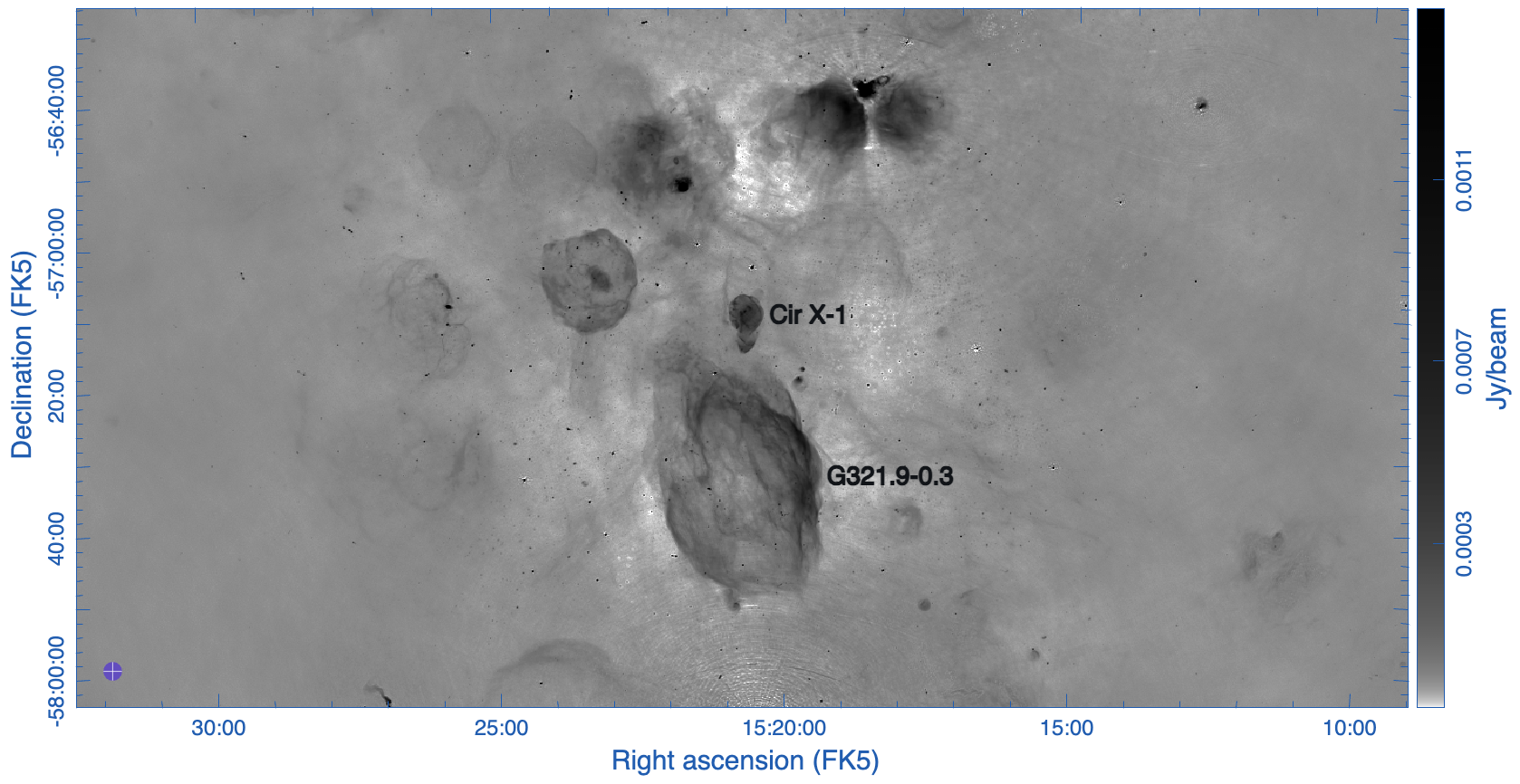}
    \caption{The MeerKAT L band image of the Cir X-1, Africa nebula and surrounding wide field with a primary beam FWHM $\sim$66\arcmin (1.1 $^{\circ}$), constructed with 30 epochs of data, we label Cir X-1 and the formally assumed supernova remnant of origin G321.9-0.3.}
    \label{fig:roughIntensity}
\end{figure*}

\begin{figure*}
    \centering
    \includegraphics[scale=0.28]{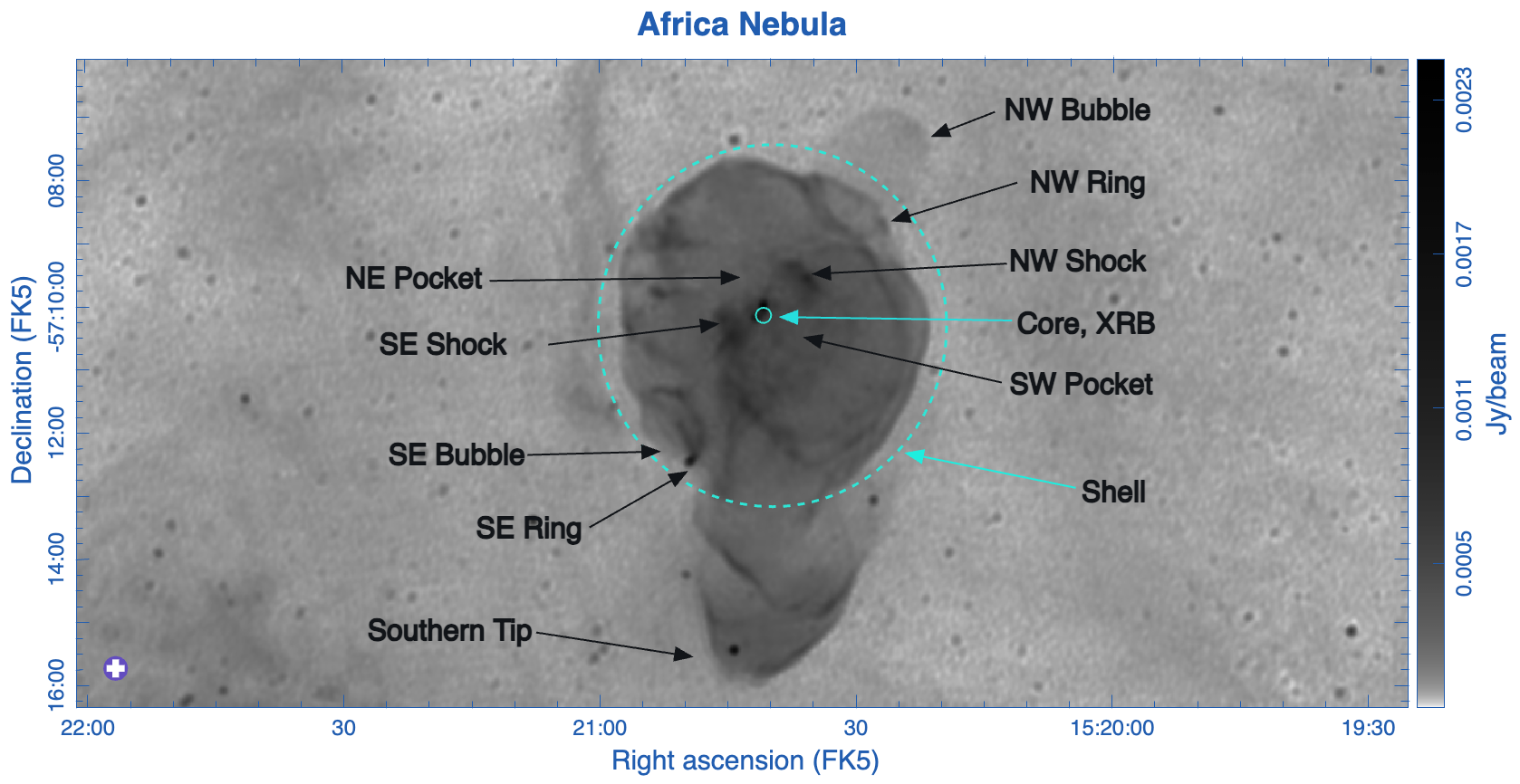}
    \caption{The zoomed-in version of Fig. 1. Highlighting Cir X-1, the Africa nebula and the newly revealed jet-punched bubbles (labelled NW and SE Bubble). We identify and label the core, XRB, the jet shock fronts, rings, pockets (more distinguishable as the spectrally flat regions in line with the core seen in the spectral index map Fig.~\ref{fig:specInd}). We  further separate the nebula into the shell region and name the extension the southern tip. }
    \label{fig:cirX1flux}
\end{figure*}

In Fig.~\ref{fig:roughIntensity} we present our deep 7.5\,hrs radio image of the Cir X-1, Africa Nebula and surrounding field, with a $3\sigma$ background rms of 38.37\,$\mu Jy/beam$. In Figures ~\ref{fig:cirX1flux}, \ref{fig:subbandIntensity} and \ref{fig:IntSpecInd}, are zoom-in versions of the deep image highlighting Cir X-1 and the surrounding Africa nebula. Fig.~\ref{fig:cirX1flux} has detailed labelling. We label the central source as `Core, XRB' and the surrounding structure of the nebula as `shell'. Within the shell we identify `jets/shock' which we label `NW/SE shock' (previously referred to as `caps' \citep{2010ApJ...719L.194S}). These are presumably the now slow precessing jets opposed to the fast fixed-axis jets assumed in Section \ref{sim_setup}. The region of space perpendicular to the jets is named the `pockets' (identifying the spectrally flat regions perpendicular to the jets which are seen in Fig.~\ref{fig:specInd}) and south of the shell is an extension we label the `southern tip'. Although, these features are visible in the previous radio images \citep{2019MNRAS.484.1672C}, we present an exceptionally resolved nebula. Improving the visibility of the asymmetric structures we call `rings' and revealing previously unobserved features such as those labelled  `bubbles', extending from the base of the rings. In the following, we analyse our observational results through a sub-band intensity map, a spectral index map and determine the minimum energy present in the NW bubble and the nebula (only the region within the shell). Subsequently, we simulate the structure of the nebula and investigate the jet parameters that may have formed the bubbles we observe in the radio image (see Section \ref{num_simulation} and \ref{sim_results}).

\subsection{Sub-band Intensity Map}\label{sec:Intensity}

\begin{figure*}
    \centering
    \includegraphics[scale=1.0]{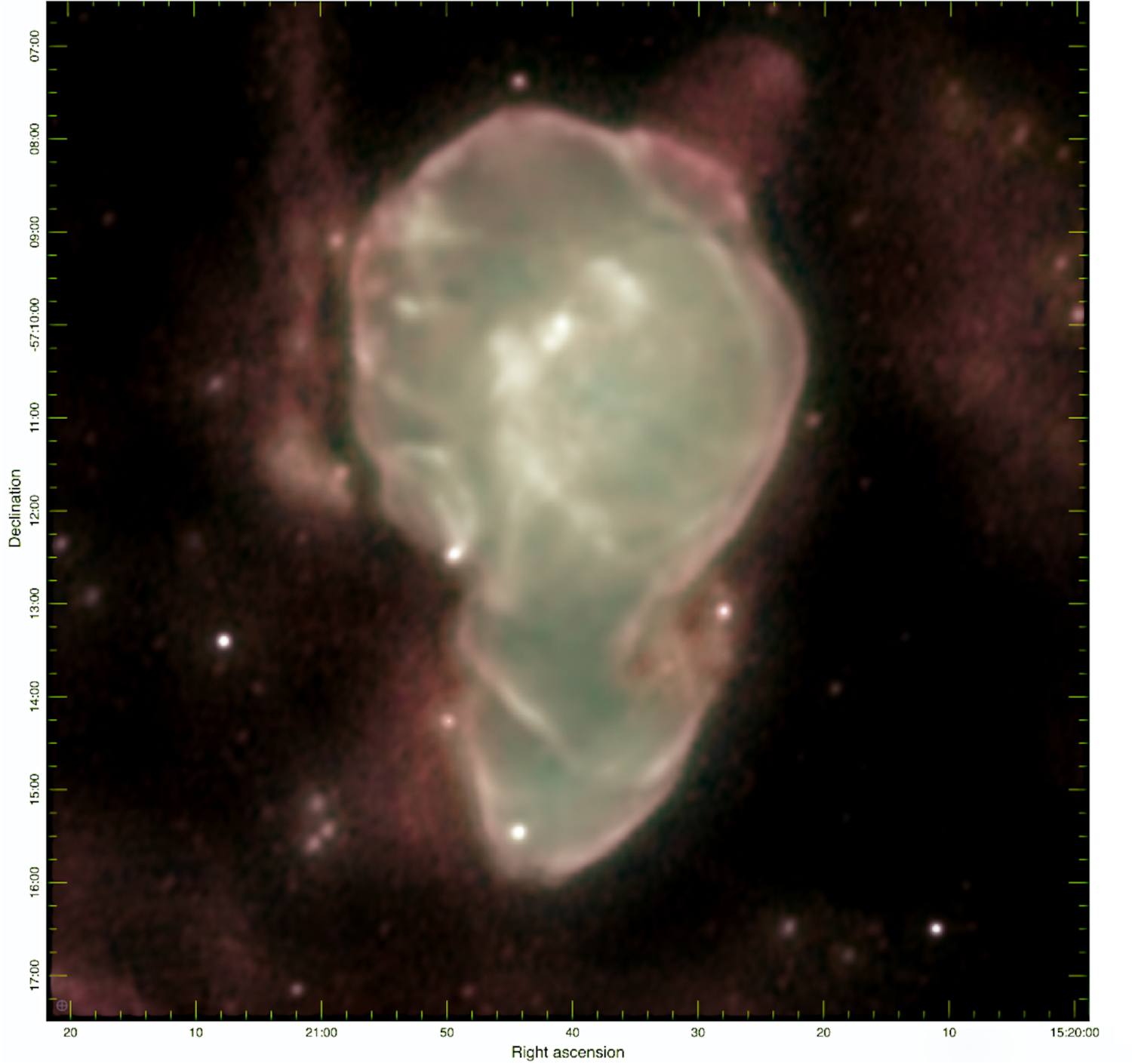}
    \caption{Sub-band Intensity Information. Colour was assigned to three sub-bands of the radio continuum observations, which were subsequently combined and adjusted as described in \protect\cite{2017IJMPD..2630010E}. The total intensity wide-band image was assigned red. The central region of the nebula was masked in the wide-band image in order to expose the colour combined sub-band image when their layers in GIMP are blended. See the text for more detail. }
    \label{fig:subbandIntensity}
\end{figure*}

In order to illustrate the features of the nebula with some indication of which frequencies dominate at which location, we constructed the composite image displayed in Fig.~\ref{fig:subbandIntensity}. Although the full frequency range of the radio continuum observation consists of 8 sub-bands, we only selected the three sub-band FITS files (without primary beam correction; 1.016 GHz, 1.230 GHz and 1.444 GHz) that were the least impacted by striations in their background noise levels. Using the CARTA visualization software\footnote{https://cartavis.org/}, we adjusted the intensities of these data using limits on the histogram of values along with logarithmic and non-linear stretches. The grayscale output png files were input into the Gnu Image Manipulation Package (GIMP)\footnote{https://www.gimp.org/}.

 Investigation of the primary beam corrected sub-bands showed that the 
  NW bubble appears intermittently between bands, depending on the signal-to-noise ratio.  This bubble is delineated most strongly in the sub-band centred at 1.016 GHz. To highlight this information, we duplicated a segment of this sub-band image capturing the bubble and blended this bubble image with the 1.016 GHz field of view.

 We subsequently assigned colours: the bubble-enhanced sub-band centred at 1.016 GHz was assigned rose, the sub-band at 1.230 GHz yellow-green, and 1.444 GHz blue-green. These colour images were intensity adjusted and combined, as described in \protect\cite{2017IJMPD..2630010E}, into a resultant colour `combined sub-band' image. 
 Next the total intensity, full-frequency (i.e. wide-band) image was assigned red. Then the inner region of the full-length of the nebula was masked in the wide-band image using GIMP's `layer mask' functionality. This exposed the coloured combined-sub-band image (described above) when the layers containing the wide-band and sub-band images 
 were in turn combined. 
Finally, masks were also employed to adjust the colour and darkness of the surrounding, non-nebular region as well as the white balance of the point sources.

\subsection{Spectral Index Map}\label{spec_index_map}
\begin{figure*}
     \begin{center}
         \begin{tabular}{cc}
     
         \subfloat{\includegraphics[width=0.5\textwidth]{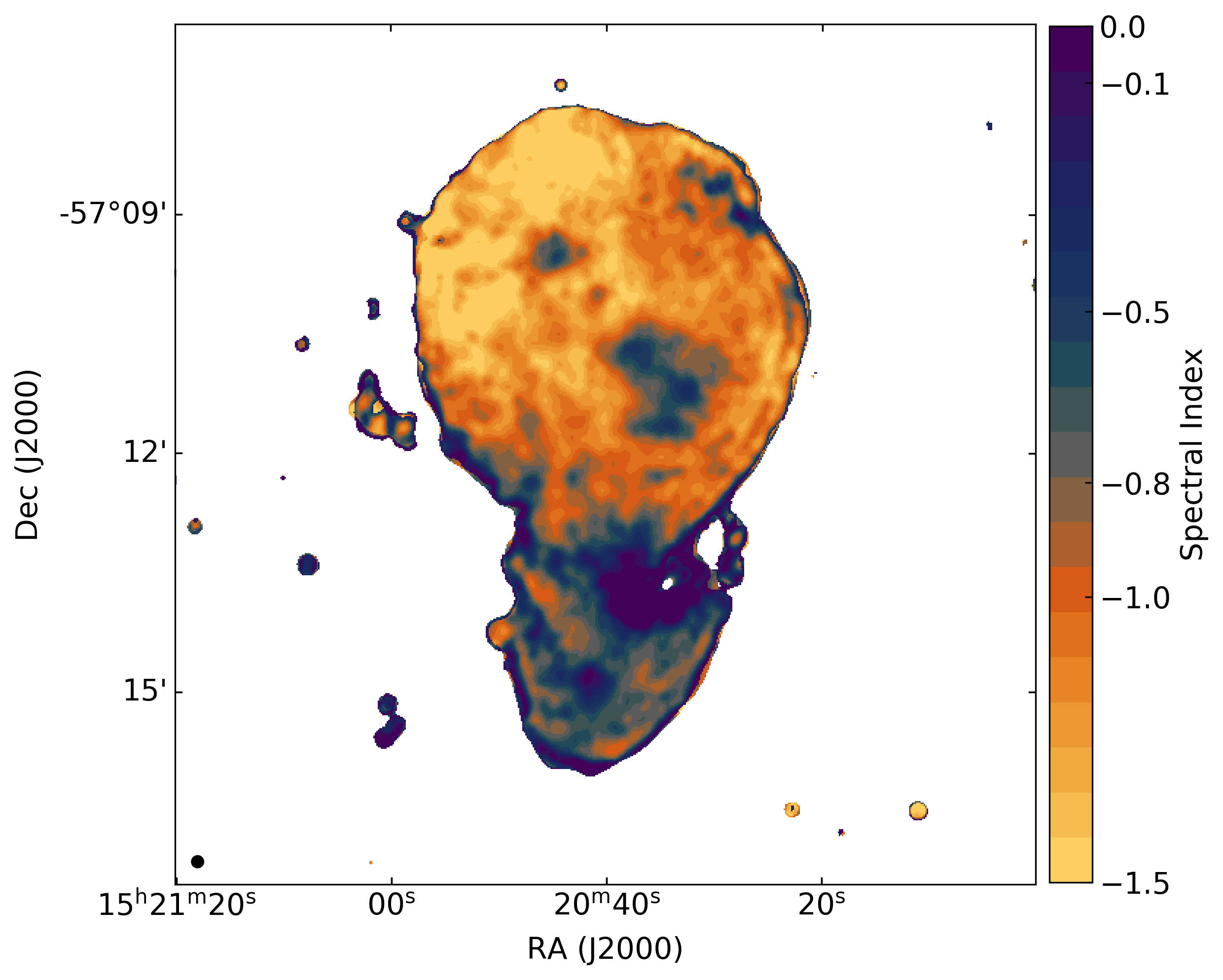}} & \subfloat{\includegraphics[width=0.5\textwidth]{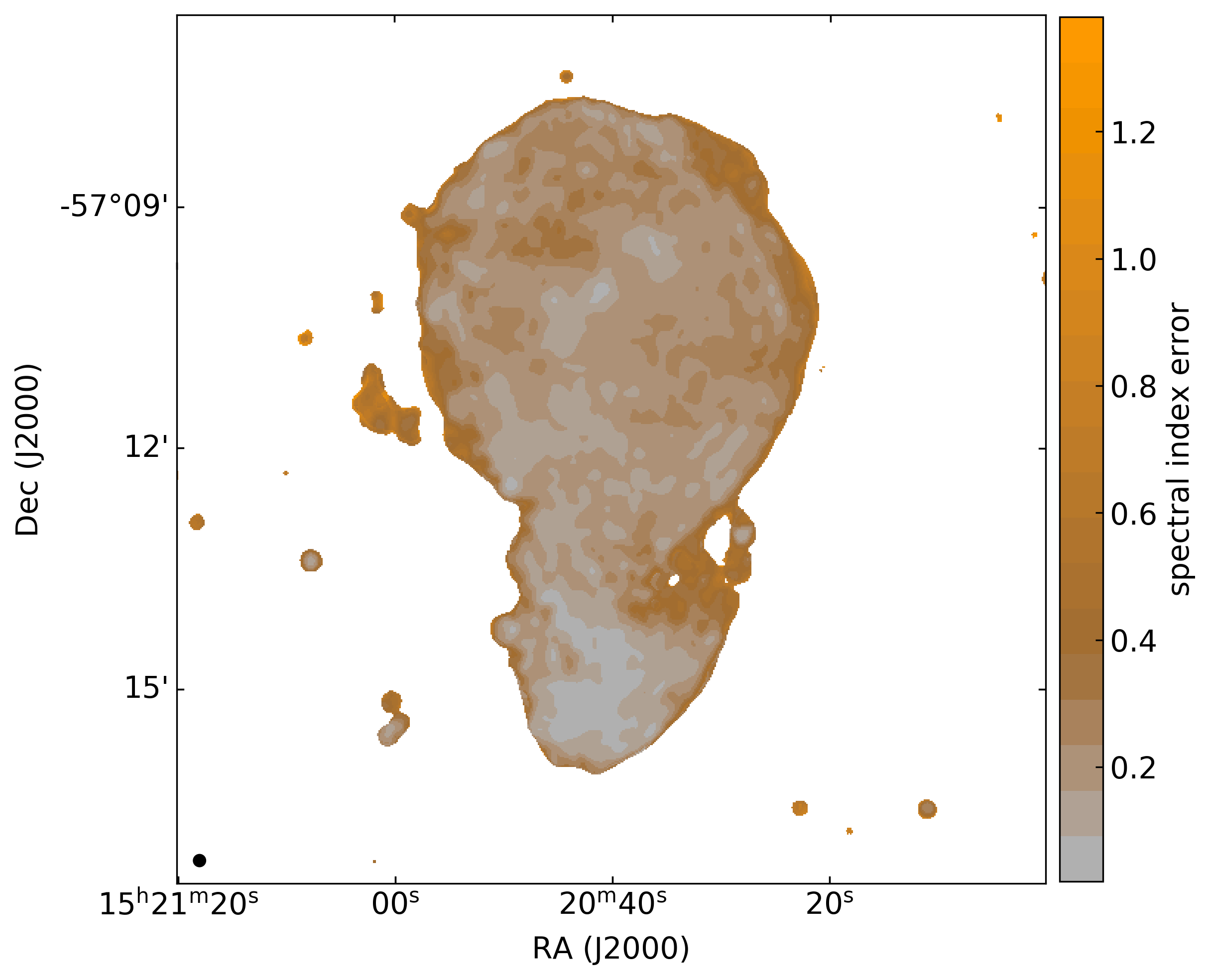}}
        
     
       
       \end{tabular}
       \end{center}
        \caption{Left panel: Spectral Index Map: Values more positive than -0.1 are conventionally referred to as flat, though these values may also be generated by noise. Values more negative than -0.8, where this colour map has a greyish divergent point, are conventionally referred to as steep. Values more negative than -1.5 are assigned yellow. The perception-based colour tables used in these maps, along with tutorials for applying them, were produced by \protect\cite{2024ascl.soft01005E} (find at https://github.com/mlarichardson/CosmosCanvas). Right panel: Spectral Index Error Map: Grey is associated with low error. Most regions in the left panel around the edge of the nebula are purple-ish due to this. However there are dark cyan through purple regions interior to the nebula with low noise that therefore have flat spectral slopes. Note that the uncertainty on the alpha values of the central point source and jet-lobes is low.} 
         \label{fig:specInd} 
\end{figure*}

We measure the global spectral index in the shell of the nebula to be -0.49 +/- 0.19. However, as seen in left panel of Fig.~\ref{fig:specInd} (the spectral index map), the shell mainly consists of values more negative than -0.8 but the global value is affected by some regions that are much less negative. Perception-based colour schemes \citep{2024ascl.soft01005E}, used in left panel of Fig.~\ref{fig:specInd}, distinguish the conventionally `steep' slopes (spectral index more negative than -0.8) in the Spectral Energy Distribution (SED) from `flat' slopes and/or noise (spectral index more positive than -0.1). The edges of the full nebula, including edges of the southern tip, are flat, which could be due to the larger uncertainties in these regions of the map (see spectral index error map in right panel of Fig.~\ref{fig:specInd}).

The centres of the shell's pocket regions are also flat, 
and, the uncertainty is lower in these regions (particularly the SW pocket). 
The bubbles, including the upper edge of the NW bubble 
are too faint to have measurable spectral indices.
We propose that this low luminosity is because they are less dense regions. In the spectral index map (Fig.~\ref{fig:specInd} left panel), the jet emission (in shock in Fig.~\ref{fig:cirX1flux}) appears to be steep (more negative than -0.8). 

Even steeper are the regions beyond the centres of the pockets contained within the shell (spectral index -1.0 to -1.5). As with other synchrotron nebula from supernova remnants to radio galaxies, the steep spectra are indicative of optically thin synchrotron emission from electrons which may have diffused from their acceleration sites. The flatter regions are likely closer to the sites of particle acceleration and/or have some optical depth.

\subsection{Minimum energy}
We determine the minimum energy for the protruding NW bubble and the nebula. The measured lengths are 1.68\arcmin\ for the NW bubble and 5.83\arcmin\ for the nebula. Using a distance of 9.4\,kpc, we determine the source size R : 7.09$\times10^{18}$\,cm for the NW bubble and 2.46$\times10^{19}$\,cm for the nebula. The minimum energy equipartition equation\footnote{see more detail in \url{https://github.com/robfender/ThunderBooks}} 
is used when the luminosity and the size of the source are measurable. Therefore, invoking Eq. (\ref{minimum_energy}), where $\eta$ is the energy associated with the protons that accompany the electron and is assumed 1 for the minimum energy condition. The filling factor (f) is assumed to be 1 and the luminosity $L$ is given by Eq. (\ref{luminosity}) for MeerKAT frequency range 909\,MHz - 1.65\,GHz. For the NW bubble $L = 1\times 10^{30}$\,erg/s and for the nebula $L =7\times 10^{31}$\,erg/s, we also determine $c_{12} = 3\times 10^7$.

\begin{equation}\label{minimum_energy}
    E_{\rm{min}} = 0.92\eta^{\frac{4}{7}}c_{12}^{\frac{4}{7}}f^{\frac{3}{7}}R^{\frac{9}{7}}L^{\frac{4}{7}}
\end{equation}

\begin{equation}\label{luminosity}
    L = 4\pi D^2F_{\nu}(\nu_2 - \nu_1)
\end{equation}
 The minimum energy that is measured is therefore 4$\times10^{45}$\,erg and 2$\times10^{47}$\,erg for the NW bubble and the nebula respectively. These values are compared to the energy estimates determined from the numerical simulations in Section (\ref{sim_setup}).
 
\section{Numerical Simulation of the Cir X-1 and Africa Nebula system}\label{num_simulation}

The deep radio images presented in this work unveil previously unseen features in this system, most notably a large and small bubble (we label NW and SE bubble respectively) symmetrically protruding out of the nebula along the axis of the known jets. The coincidence of these bubbles along the jet axis suggests that past jet activity could indeed be responsible for or at least linked with the formation of these bubbles. We hypothesize that these bubbles are punctures in the supernova remnant caused by the jet launched by Cir X-1: the jet is launched some time post-SN explosion, the jets catch up with the SN remnant, and then they interact with the remnant such that they blow the observed bubbles which we see today.

Although jet activity is visible in this image (presented and discussed in detail in \citet{2019MNRAS.484.1672C} and Cowie et al 2025. in prep), the apparent precession angle of these jets simply gauged from the width of their termination shocks appears too wide to be responsible for the comparatively narrow puncture at the base of the bubbles. Additionally, the jets appear to produce termination shocks within the nebula (labelled as shocks in Fig.~\ref{fig:cirX1flux}), rather than at the surface of the nebula. We therefore propose that there exists at least two jet modes in Cir X-1.
Cowie et al. (2025, in prep) propose that the jets responsible for the inner shocks visible in the radio image are relatively slow ($v_{\rm jet}\sim0.3$c), 
while we propose a set of fixed-axis fast jets ($v_{\rm jet}\sim$ c) launched at an earlier time
are responsible for the radio bubbles we see today. 

To test this hypothesis, as a proof of concept, we run numerical simulations representative of the basic components we observe in the system (a supernova explosion, and a set of symmetric jets launched from the core) to try to reproduce the observed morphology. 

\subsection{Numerical Methods}

The simulations in this paper are relativistic hydrodynamic (RHD) simulations run using the \texttt{PLUTO} code \citep{Mignone2007}. The code computes the numerical solutions to the RHD system of equations, evolving the conservative set of variables

\begin{center}
\begin{math}
    \boldsymbol{U} = (D,m_1,m_2,m_3,{\cal E}_t)^T
\end{math}
\end{center}

\noindent with laboratory mass density $D$, momenta $m_{1,2,3}$ and total energy ${\cal E}_t$ densities,  according to the conservation equations

\begin{center}
\begin{math}
\frac{\partial}{\partial t}\begin{pmatrix}
    D \\ m \\ {\cal E}_t
\end{pmatrix} + \nabla \cdot \begin{pmatrix}
    D\boldsymbol{v} \\ \boldsymbol{mv} + P {\cal I} \\ \boldsymbol{m}
\end{pmatrix}^T = \begin{pmatrix}
    0 \\ 0 \\ 0
\end{pmatrix}
\end{math}
\end{center}

\noindent with velocity $\boldsymbol{v}$, thermal pressure $P$ and identity matrix ${\cal I}$. 

At each timestep, the code converts these conservative variables to primitive variables in the lab-frame (rest-mass density $\rho$, three-velocity $\boldsymbol{v}$ and pressure $p$) which make up the outputs of the simulation, as well as the initial and boundary conditions. Lastly, we employ the Taub-Matthews equation of state \citep{taub1948relativistic,mignone_equation_2007} which describes a relativistic perfect gas as a function of temperature, and is therefore flexible to both relativistic and non-relativistic regimes. 

We assume a 2D cylindrical geometry in the simulations, which lends itself well to the observed geometry of the nebula shell and bubbles, neglecting the southern tip feature which we are not aiming to reproduce in the scope of this work. The simulation is aligned such that the jet runs along the axisymmetric axis of the simulations, and the core is in the centre of the simulation. 

The domain is a static grid split into $1500\times3000$ computational cells. This constitutes the 2D grid which is rotated completely about its longest axis, making each cell square toroidal 
in shape and occupying a pseudo-3D volume. The boundaries are set to be outflowing, with the exception of the axisymmetric boundary the simulation is rotated about. 

The simulation runs in non-dimensional `code' units which are translated to physical units with time and length scales appropriate for the system. These scaling units are defined as: unit density $\rho_0=1.673\times 10^{-24}$\,$\text{g cm}^{-3}$, unit length $L_0=3\times10^{16}$ cm, and unit velocity $v_0=c$, making units of time $t_0=3.173\times10^{-2}$ years, or $1.001\times10^{6}$ seconds. Each simulation computational cell is one code unit across, meaning the domain has a physical resolution of $15\times30\times 10^{18}$cm. The resolution was selected such that the morphology of the simulations remained consistent if the resolution was increased, but did not pose excessive computational costs.

In terms of numerical integration, the code offers several choices for spatial reconstruction and time-stepping algorithms. 
We use a linear reconstruction of primitive variables as the spatial order of integration, and a $2^{\rm nd}$ order total variation diminishing (TVD) Runge-Kutta scheme to evolve timesteps. To compute the flux between cells, we use the Harten–Lax–van Leer contact (HLLC) Riemann solver \citep{mignone_hllc_2005}, along with the least diffusive flux limiter available in \texttt{PLUTO} (monotized central difference). Lastly, we enable additional dissipation in the proximity of strong shocks (`shock-flattening') with a multi-dimensional strategy. 

\subsection{Simulation Setup}\label{sim_setup}

We construct a simulation to reproduce a basic supernova explosion, and a relativistic jet launched from the position of the core. We do so with methods similar to those presented in \cite{goodall2011}, a similar study of an `older-sibling' analogous system, SS433.

The HMXRB SS433 exhibits remarkably similar properties to Cir X-1, most notably its positioning within its assumed natal supernova remnant (the W50 nebula) which also shows asymmetric bubble-like features along its jet axes. Although SS433 is a comparatively older example of this phenomenon (Likely $10^4$ to $<10^5$ years old), many of the initial simulation conditions that \cite{goodall2011} used to explore the effects of the inner jets on the morphology of the surrounding nebula translate well to this work. We do not, however, model a precessing jet like in \cite{goodall2011} but instead launch the jet along a fixed axis, which we expect for reasons outlined at the beginning of this section.

\subsubsection{Supernova explosion}

To model the initial natal supernova explosion, we simulate a Sedov-Taylor \citep{taylor1950,sedov_examples_1958} spherical blast wave at the centre of the simulation volume. This Sedov expansion is a standard approximation of a supernova explosion once the swept up mass exceeds the initial supernova ejecta mass, immediately following the initial kinetic `free-expansion' phase during which the ejecta moves at a constant ejection velocity and expands more quickly than the subsequent Sedov phase. The properties of the emitting gas in the supernova remnant are consistent with a type IIP supernova explosion with progenitor mass ranging from $\sim8$ to $\sim25 M_{\odot}$ \citep{chevalier2005}, and thus we make a conservative estimate of about $5M_{\odot}$ of ejected mass. With this mass, the radius at which the Sedov phase of expansion begins is $\sim3.6$ pc for a spherical explosion. The current radius of Cir X-1 at $9.4$ kpc is roughly 8 pc across and thus is likely currently in the Sedov phase, and was only dominated by the free-expansion phase for the first several hundred years of its evolution, a fraction of this simulation time. Simulating the free-expansion phase would accelerate the growth of the supernova, as it is more rapid in this phase, and would produce a more realistic mass distribution, but would require a much higher resolution simulation that is beyond the scope of this investigation.

To simulate the supernova explosion, we initiate a sphere with pressure
\begin{equation}
    p_{\rm SN}=\frac{(\gamma-1)E_{\rm SN}}{V}
\end{equation}
where $E_{\rm SN}=3.0\times10^{50}$ ergs, we assume a non-relativistic adiabatic index $\gamma=5/3$, and $V$ is the volume of the initial explosion which we set to be spherical with radius $R_{\rm SN}=3$ simulation pixels, equivalent to $9\times10^{16}$cm ($\sim$0.03 pc). Inside the explosion radius, we set the density to be equal to the ambient density. 

\subsubsection{ISM and Galactic density profile}
Outside of the explosion radius, the pressure is set as
\begin{equation}
    p_{0}=\frac{c_{s,0}^2 \rho_{0}}{\gamma}
\end{equation}
where the sound speed in the ambient medium is $c_{s,0}=9\times10^{5}\rm{cm}\,\rm{s}^{-1}$, the density is set at $\rho_{0}=1.673\times10^{-24}\rm{g}\,\rm{cm}^{-3}$, equivalent to $n_0=1\,\rm{cm}^{-3}$ for a fluid of protons. This pressure is constant everywhere outside of the explosion radius to ensure stability in the surrounding ISM. 

The density in the surrounding ISM is not constant, however, as Cir X-1 sits very close to the Galactic plane and therefore likely sits in a region of high density gradient (again, much like its counterpart SS433). To simulate this, we again adopt methods from \cite{goodall2011} and employ an exponential density profile of the Galactic plane as described by \cite{dehnen1998mass}, adopted for the position of Cir X-1 with respect to the galactic plane and normalized such that the number density at the centre of the simulation (the position of the XRB) is unity. We measure the position of Cir X-1 relative to the Galactic centre as $R_\star=5.8$ kpc and $z_\star=6.15$ pc. The jet axis, which we infer to be aligned with the bubbles on either side of the nebula, is perpendicular to the galactic plane, an advantageous feature as it therefore allows us to maintain an axisymmetric geometry in our simulation. Lastly, we inject random Gaussian perturbations in the surrounding ISM of the scale $10^{-4}$.

\subsubsection{Jet injection}

The jet in this simulation is launched after the initial supernova explosion from the centre of the simulation volume and explosion epicentre. We inject the jet as a relativistic flow through a nozzle which is set as an internal boundary, meaning the cells which inject the jet are not evolved according to the fluid equations along with the rest of the grid. Symmetric jets are injected along the z-axis in the positive and negative direction. The jet nozzle spans 3 pixels in the x-direction and 1 pixel in the z-direction, and therefore the injection radius is $9\times10^{16}$cm -- this being the smallest area for a numerically stable jet, although we note that this is much larger than the orbital separation of the binary system $\sim10^{12}$cm \citep{tauris1999}. We set the pressure in the jet as $7.5\times10^{-10}\,\rm{g}\,\rm{cm}^{-1}\,\rm{s}^{-2}$, such that the jet is hot with respect to the ISM, but depending on the pressure within the supernova at the time that the jet is launched, is at times slightly cooler than its surrounding fluid. Jet pressure is kept constant in all simulations as an independent variable.

The main goal of the simulation is to reproduce the morphology of the jet-nebula interaction. As such, we choose to only vary the parameters of the jet, as we assume that this has the largest influence on the resultant morphology, and they are currently poorly understood. The main jet parameters we vary are:

\begin{minipage}[t]{0.45\textwidth}
\begin{itemize}[align=left,labelsep=1ex]
\raggedright
    \item[$P_{\rm jet}$:] the jet power, mainly through varying the jet Lorentz \makebox[0.7cm]{\hfill}factor $\Gamma$, although in some simulations we test two jet \makebox[0.7cm]{\hfill}densities $\rho_{\rm jet}$
    \item[$t_{\rm jet}$:] the jet initial launch time, measured in years since \makebox[0.7cm]{\hfill}supernova
    \item[$\Delta t_{\rm jet}$:] the jet injection duration, which contributes to the total \makebox[0.7cm]{\hfill}injected energy $E_{\rm jet}$

\end{itemize}
\end{minipage}
\kern.1\textwidth

These all have a significant effect on the final morphology of the jet-nebula interaction site, and varying these allows us to understand how certain morphological features of Cir X-1 may have arisen. 

\subsection{Simulation analysis}

To compare the simulation directly with the radio data presented here in this work, we estimate a pseudo-emissivity from the simulation data which serves to replicate optically thin synchrotron emission following the prescription outlined by \cite{hardcastle2013numerical}. Based on the assumption of minimum energy, where there is equipartition between the energy in magnetic fields and electrons, the comoving\footnote{ we denote all comoving quantities with the prime $'$ notation.} synchrotron emissivity can be expressed as

\begin{equation}\label{emissivity_comoving}
    j'_{\nu'}= A ( p ,\nu',\gamma,\eta)(\kappa P)^{(p+5)/4}
\end{equation}
\noindent where $P$ is the pressure in the simulation, $p$ is the spectral index, $A( p ,\nu',\gamma_{1,2},\eta)$ is the constant prefactor for all simulation cells \citep[see][]{hardcastle2013numerical} which depends on the spectral index, the comoving observing frequency $\nu'$, the minimum and maxium Lorentz factors of the electron distribution $\gamma_1=10$ and $\gamma_2=10^5$, and $\eta=0.75$ is the partitioning factor between energy in the magnetic fields and electrons. This is related to Equation \ref{minimum_energy} used to calculate the minimum energy from the observed emission, but includes an extra term $\kappa$. This term provides a simplistic treatment for particle acceleration, delegating how much of the internal pressure is stored in relativistic electrons and magnetic fields. We set $\kappa=0.1$, which roughly translates to a particle acceleration efficiency of $\sim5\%$. This is a reasonable estimate, as we know roughly $10\%$ of shock power from supernova remnants is thought to be funnelled into cosmic ray acceleration \citep{helder2009,bell013,morlino2013}. 

The emissivity in Equation \ref{emissivity_comoving} is a comoving quantity, and relativistic effects should be incorporated when calculating the flux in the frame of the observer. This flux is given as 
\begin{equation}
    F_{\nu}=\frac{\delta^{2-\alpha}}{d_{L}^2}\int j'_{\nu}dV
\end{equation}
where $V$ is the lab-frame volume\footnote{the \texttt{PLUTO} code returns primitive variables (in the comoving frame) but the grid is defined in the lab frame, hence why we integrate over the lab frame volume.} over which the emission is integrated, $\alpha$ (defined as $p$ in Equation \ref{emissivity_comoving}) is the spectral index. While we measure a global spectral index of $\alpha$  $\sim -0.5$ (Section \ref{spec_index_map}), this is affected by optically thicker pockets. Since we are interested in simulating the optically thin emission, we adopt a spectral index of $\alpha=-1$, which is representative of this emission in the shell (Fig.~\ref{fig:specInd}). $d_L$ Is the distance to the source, and $\nu=1.28$GHz is the lab-frame observing frequency. The Doppler factor $\delta$ is given as

\begin{equation}
    \delta=\Gamma^{-1}(1-\beta\cos{\theta})^{-1}
\end{equation}
where $\theta$ is the angle of the velocity vector to the line of sight, $\beta$ is the speed of the fluid in units of c, and $\Gamma$ is the Lorentz factor. 

We note that when calculating the minimum energy of the nebula and bubbles from the radio data in Section \ref{minimum_energy} we do not make any relativistic corrections. This is predominantly because we do not know the velocities present in the system and therefore can not correct for them, and secondly, as we infer and discuss later given the simulation results, we deem it unlikely that there is a presently active jet producing the bubbles and therefore the Lorentz factor of the observed fluid today is likely not relativistic.

Lastly, if we assume that the emission is optically thin, we are able to make maps of the integrated emission across the 3D volume. To do this, we rotate the 2D cylindrical simulation frames about their z-axis and interpolate them onto a 3D Cartesian grid. This data is then integrated along the line of sight using an in-house the ray-tracing algorithm \texttt{DART} to produce images which we then use to compare to the real radio image.

\section{Simulation Results}\label{sim_results}

We ran several test simulations, varying jet power and duty cycle within a confined parameter space, in order to gauge how the jet-nebula morphology changes as a result. These simulations non-comprehensively spanned the parameters listed in Table \ref{Tab:jetparams}. Although we did not test all permutations, simulations were run in batches with control and independent variables, after which several test simulations of interest were pursued. As a result of this process, we present a fiducial simulation, presented in Figures \ref{fig:sim_timesteps}, \ref{fig:raytrace} and \ref{fig:multisim_comparison}, which is best able to reproduce the features observed in the radio image Figure \ref{fig:cirX1flux}. The fiducial simulation parameters are shown in Table \ref{Tab:fidsim_params}, as well as the parameters for two select comparison simulations which we present in Figure \ref{fig:multisim_comparison}.

\begin{table*}
\centering
\captionof{table}{Jet parameter values adopted in test simulations. A non-comprehensive number of permutations were tested.}\label{Tab:jetparams}
\begin{tabular}{rll}
\hline
{Parameter} & {Values} & Description\\ \hline
 $\Gamma_{\rm jet}:$ & 3, 5, 7,  10  & Lorentz factor of the jet \\
$n_{\rm jet}:$ & $7.7\times10^{-5}$ , $10^{-6}$ & Jet density (cm$^{-3}$)\\ 
$\Delta t_{\rm jet}:$& 500, 550, 800, 1000, 1500 & Jet duration (years)\\ 
 $t_{\rm jet}:$& 50, 100, 400, 1000, 2000, 2500, 3000  & Jet initial launch time (years) \\  

\hline
\end{tabular}
\end{table*}

\begin{table*}
\centering
\captionof{table}{Simulation parameters of the fiducial simulation, as well as two select comparison simulations which we present in Figure \ref{fig:multisim_comparison}.}\label{Tab:fidsim_params}
\begin{tabular}{lllll}
\hline
 {Simulation name} & {$\Gamma_{\rm jet}$} & $\Delta t_{\rm jet}$ & $t_{\rm jet}$ & $n_{\rm jet}$ \\ \hline
 Fiducial simulation & 7  & 550 years & 50 years &  $7.7\times10^{-5}$ cm$^{-3}$ \\
Increase $t_{\rm jet}$ & 7 & 800 years & 2500 years &  $7.7\times10^{-5}$ cm$^{-3}$\\ 
Increase $\Delta t_{\rm jet}$& 7 & 1500 years & 50 years &  $7.7\times10^{-5}$ cm$^{-3}$\\

\hline
\end{tabular}
\end{table*}

\subsection{A fiducial simulation}

\begin{figure*}
    \centering
    \includegraphics[width=1.0\linewidth]{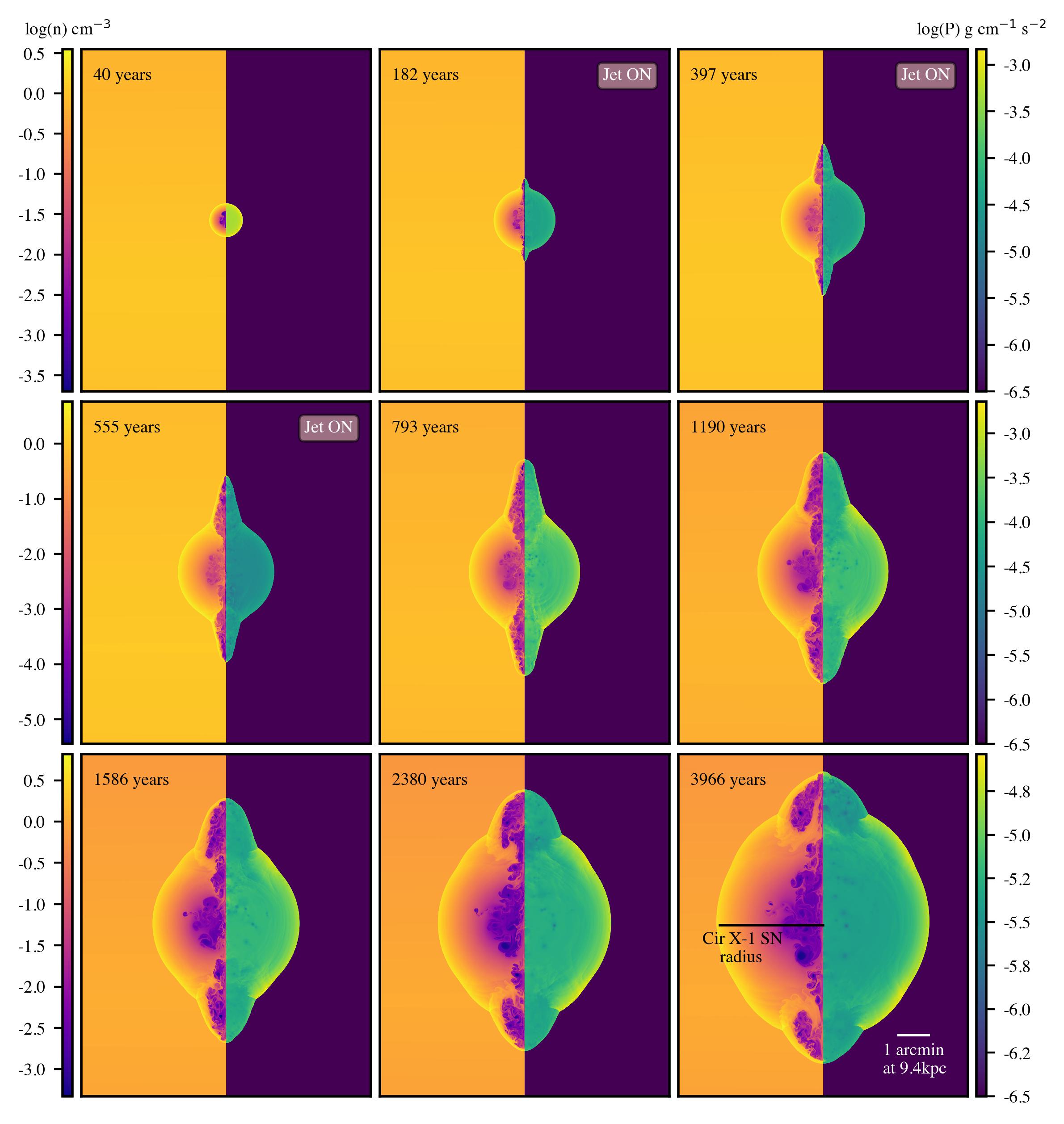}
    \caption{Various timesteps of the fiducial simulation, with the time elapsed since initial supernova indicated in the top-left. Each panel shows the log of the number density on the left-hand side, and the log of the pressure on the right. Each row shares the same colourbar axes on the left (number density; lower densities assigned darkest colours, e.g. purple) and right (pressure; lower pressures assigned darkest colours, e.g. blue and purple) of the row. In the final timestep, we indicate the measured radius of the Cir X-1 supernova remnant as well as the angular scale of the simulation at the same distance as Cir X-1. We indicate in the top-right corner of each panel if the jet is active during that timestep, where the jet is launched between 50-600 years in this simulation.}
    \label{fig:sim_timesteps}
\end{figure*}

We are successfully able to reproduce a supernova remnant of the correct size within the age estimate, along with a set of inflated bubbles along the jet axis. The formation of these bubbles, due to the interaction between the jet and the supernova remnant, can be seen clearly in Figure \ref{fig:sim_timesteps}. 

The supernova nebula in the simulation prior to the jet launching is initially spherical. At 50 years, the jet is launched, and shortly after we begin to see protrusions on either side of the nebula where the jet has propagated past the supernova shock radius. While the jet is still on (see panels at 182, 397, and 555 years in Figure \ref{fig:sim_timesteps}) the jet produces a pointed, conical feature as it drives into the ISM. At this time, these features are in rough pressure equilibrium with the nebula. After the jet turns off at 600 years, the jet is no longer driving the bubble features into the ISM and instead they begin to expand and cool -- this results in the previously conical bubbles becoming more rounded at their tip and expanding outwards (see, in particular, the progression from 555-1190 years along the middle row in Figure \ref{fig:sim_timesteps}).  We also begin the see the bubbles become cooler than the inner nebula over this time. The bubbles continue to expand adiabatically, as well as with the supernova explosion itself, and the base of the bubble where the jet had essentially punched a hole into the remnant widens with the expansion of the supernova, resulting in the flattened morphology of the bubble we see at the end of the simulation. 
\\

This fiducial simulation presents a scenario where a powerful jet, approximately $\sim35 L_{\rm Edd}$ for a $1.4M_{\odot}$ neutron star, is launched early in the evolution of the supernova over a relatively short amount of time. This is consistent with the fact that we do not observe any other features evidencing a recently active powerful jet. Additionally, we observe what seems to be a slower precessing jet unimpeded by any other faster jet activity, such that the slow jet must be launched after the fast jet (or at the very least along another axis).

\begin{figure*}
    \centering
    \includegraphics[width=0.85\linewidth]{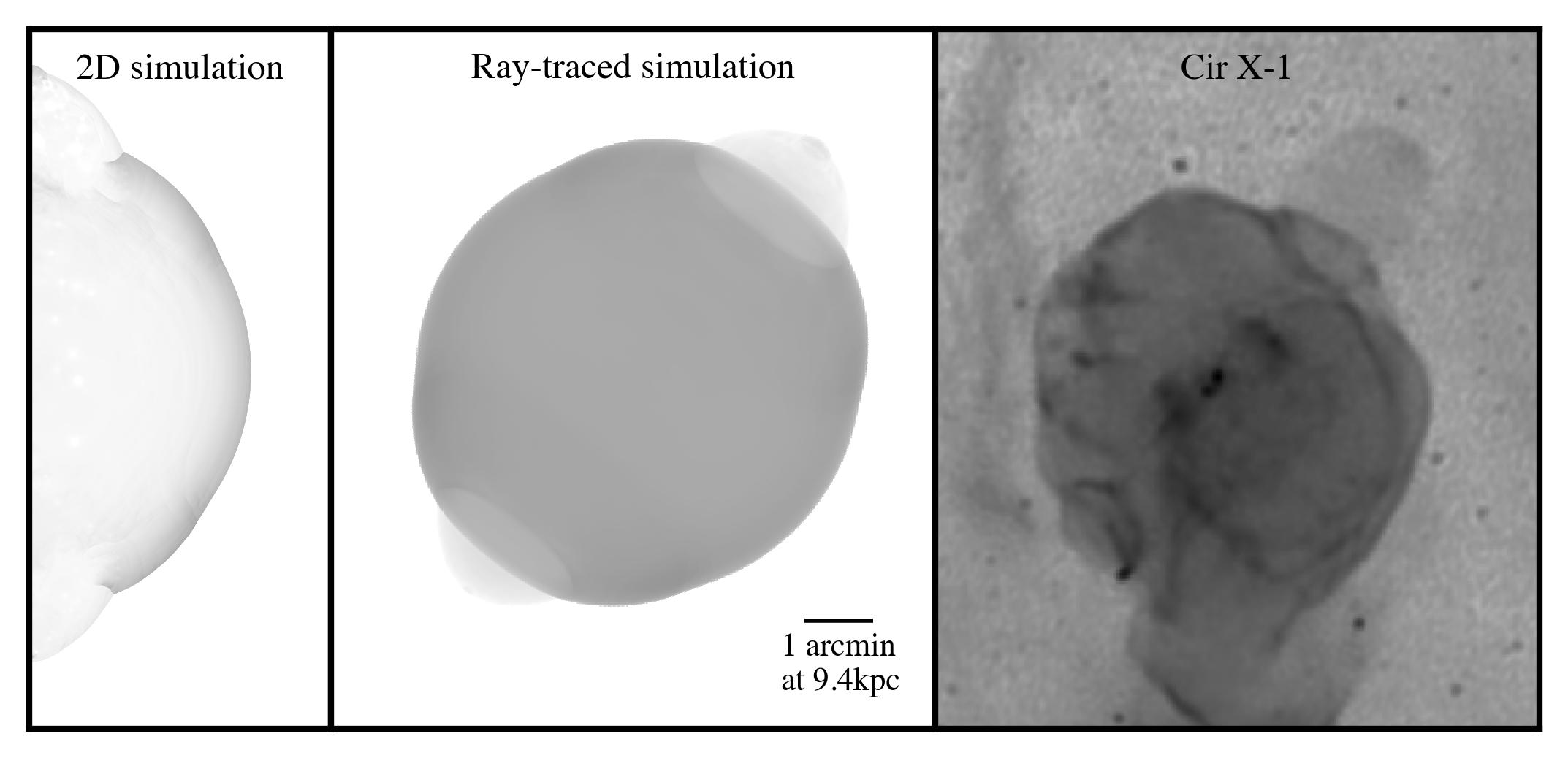}
    \caption{Lab-frame flux of the fiducial simulation compared to the observed radio flux. The 2D axisymmetric simulation (left) is rotated about its axis and the emissivity is integrated along a 60$^\circ$ angle to the line of sight using the in-house ray-tracing software \texttt{DART} to produce the optically thin flux image (centre). We compare this morphology to the real radio image of Cir X-1 (right). Flux is displayed in log-space, the colour scales of the images are individually and arbitrarily scaled for visual comparison, and the length scales are equal in all panels. We indicate in the central panel the angular scale of the simulation at the same distance as Cir X-1, which also spatially corresponds to 1 arcminute in the real radio image on the right.}
    \label{fig:raytrace}
\end{figure*}

In addition to the simulated jet producing a pair of bubbles, matching those observed in the data, we are also able to reproduce several other key features present in the radio image. Here, we compare the ray-traced flux as seen in Figure \ref{fig:raytrace} with the Cir X-1 radio image.
\\

Firstly, the 
region
where the inflated radio bubbles meet the edge of the supernova remnant in Cir X-1 is slightly enhanced in brightness, producing a ring-shaped base to the bubble -- we also 
reproduce this in the fiducial simulation. Secondly, the tip of the radio bubble is slightly brighter compared to the rest of the bubble which is uniformly illuminated, which we also observe in the simulation, most apparent in the emissivity map of the fiducial simulation in the central panel of Figure \ref{fig:multisim_comparison}. Lastly, in the radio image, the bubble itself has a strikingly smooth and round shape, especially in contrast to the complex structure seen inside of the supernova remnant. This smooth shape in the bubble is reproduced in the fiducial simulations, although we do not see the same amount of structure in the nebula itself (this is likely due to our simplistic treatment of the supernova explosion, although we do see some structure appear in other simulations which we discuss further in Section \ref{manysim_results}).

\subsection{Relations between jet parameters and radio features}\label{manysim_results}

\begin{figure*}
    \centering
    \includegraphics[width=1.\linewidth]{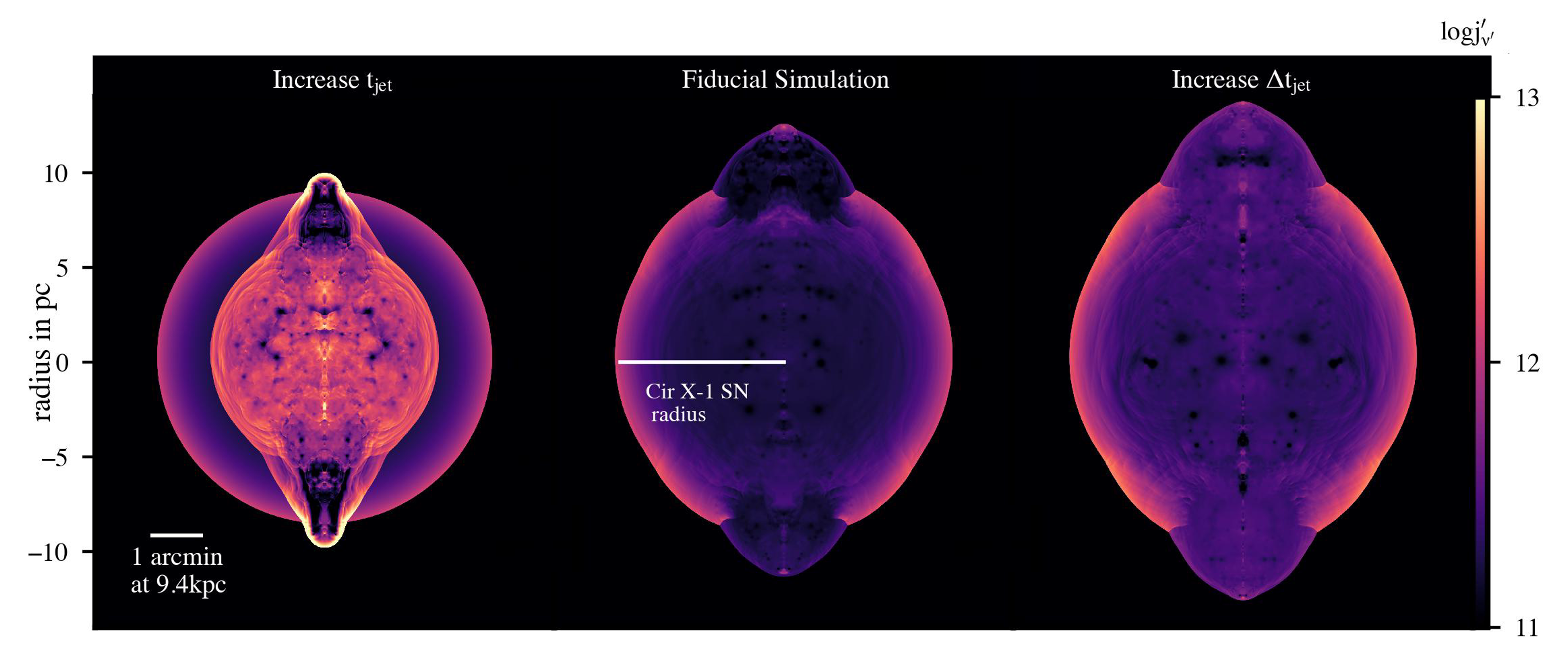}
    \caption{Comparison between the fiducial simulation (centre), a simulation where the jet is launched later in time and for slightly longer (left), and the same simulation but the jet is launched for a longer period of time (right) but starts at the same time as the fiducial. Each simulation is shown at 3966 years old. We indicate the measured radius of the Cir X-1 supernova remnant for comparison, as well as the angular scale of the simulation at the same distance as Cir X-1. Each map is showing the log of the comoving emissivity $j'_{\nu'}$ per pixel. The scale on the left of the image is the radius of the bubble with respect to the centre of the supernova in parsecs. Simulation parameters are specified in Table \ref{Tab:fidsim_params}.}
    \label{fig:multisim_comparison}
\end{figure*}

With the fiducial simulation we show as a proof of concept a plausible scenario for the formation of these bubbles. However, there are certain features in the radio image which are not reproduced well in the fiducial simulation, but are present in other test simulations (although at the expense of other key features). By running these test simulations, we are able to broadly infer which jet parameters are responsible for various features and in general make statements about how these jets affect their natal supernovae. 

Here we enumerate radio features present in the data which we have aimed to reproduce in simulations, and qualitatively discuss how the jet parameters affect the emergence of these features. 

\begin{description}
\setlength\itemsep{1em}
    \item[Size and brightness of the bubble rings.] The ring formed by the intersection of the bubble and the supernova remnant is a prominent feature for both bubbles (NW and SE) in the radio image -- the ring is not only slightly brighter than the adjacent emission from the nebula, it is also narrow, in that it occupies a relatively small surface area of the nebula. Comparing the radio image to the fiducial simulation, the rings in the simulation are wider and are slightly less edge-brightened. 
    
    Addressing the size of the ring, this is mainly because the jet in the simulation is launched relatively early in the evolution of the supernova. A hole is punched by the jet into the edge of the expanding supernova remnant shortly after launch, as seen in Figure \ref{fig:sim_timesteps}. As the remnant expands, the puncture also expands with it, producing a wide-radius ring at the end of the simulation evolution. The rings produced by jets launched at late times are narrower, but produce much smaller bubbles for the same power jet, see the left panel in Figure \ref{fig:multisim_comparison}. These late-launch simulations also show marginally more edge-brightening, but we found this varied inconsistently across simulations and likely relies on more complex interplay between the jet/nebula hydrodynamics.

    \item[Height and shape of the bubble.] One of the most striking features of the Cir X-1 bubbles is how prominent the northern bubble is: its height along the axis of the jet is almost equal to the radius of the nebula itself ($\sim$80\% of radius). In addition to this, the bubble appears to be widest at halfway between the nebula puncture and its tip, giving it the distinctive `bubble-like' appearance. We were unable to produce a simulation within our explored parameter space with this `bubble-like' shape, and all bubbles produced were instead widest at their base. Recently active jets (as seen at t=55 and 793 years in Figure \ref{fig:sim_timesteps}) produce bubbles which are triangular in shape, and after the jet is turned off the bubbles expand adiabatically to produce more rounded structures. This suggests that the jets in Cir X-1 have been switched off, or at least operating at substantially lower power, for enough time for this expansion to have taken place, as we do not observe a triangular morphology in the radio image.
    We also note that the jet requires less power to puncture the expanding supernova shell and produce a bubble of sufficient height when launched early, as opposed to late. However, even with a jet power equivalent to  $\sim35L_{\rm Edd}$ launched only 50 years after supernova (as adopted in the fiducial simulation, at the very high end of what we would expect regarding jet power from a NS), the bubble produced is still too short. Jets which were sustained for a longer $\Delta t_{\rm jet}$ produced bubbles which were taller, but also slightly wider at their base and overall less rounded, as evident in the right panel of Figure \ref{fig:multisim_comparison}.
    
    \item[Bubble asymmetry.] Another remarkable feature of the radio bubbles observed in Cir X-1 is their apparent asymmetry. To reproduce this asymmetry, we include a model of the galactic density profile in some of our tests. This profile results in the southern bubble growing in a denser local environment compared to the northern bubble, resulting in a reduced size and a flatter profile in the southern bubble. However, the extent to which they are asymmetric is not reproduced, and the simple galactic profile is not sufficient to explain the observed asymmetry. 
    \item[Ratio of bubble and supernova remnant flux density.] Although Cir X-1 is relatively well studied, the bubbles were not detected until now due to their low surface brightness. This feature, although somewhat frustrating for earlier observers, is particularly telling of their history. In simulations, the bubbles are bright compared to the expanding nebula when formed, in particular the tip of the bubble which is the working surface of the jet. Only after hundreds of years of cooling do they reach such low surface brightness. We see this in Figure \ref{fig:multisim_comparison} between the fiducial simulation and the simulation on the right launched over a longer $\Delta t_{\rm jet}$, where the latter has not had enough time to cool. This adds to the mounting evidence that the jets were launched early in the history of the remnant. Additionally a longer cooling time limits the total energy injected by the jet into the bubble and therefore implies a shorter $\Delta t_{\rm jet}$.
    \item[Features inside the supernova remnant.] The emissivity inside of the nebula in the fiducial simulation is very uniform due to the smooth pressure profile of the Sedov-Taylor expansion, and the absence of internal shocks from the jet injection. This is not the case, however, for simulations where the jet is launched later in the supernova evolution, see an example in the left panel of Figure \ref{fig:multisim_comparison}. In these simulations, the interior of the supernova has cooled and the sound speed has decreased, enabling the jet to inflate a cocoon of reheated material which produces a secondary shock within the nebula itself. This secondary shock propagates outwards and fills the space inside the nebula, creating a turbulent spherical structure brighter than the outer remnant. In the radio images, we could interpret some of the inner structure to be of similar morphology, and possibly be caused by jet activity. However, supernova remnants without jet activity have inherently complex turbulent internal structures, and it is likely that a more realistic treatment of the supernova explosion could also produce structures akin to those seen in Cir X-1.  
    
\end{description}

\begin{figure*}
    \centering
    \includegraphics[scale=0.9]{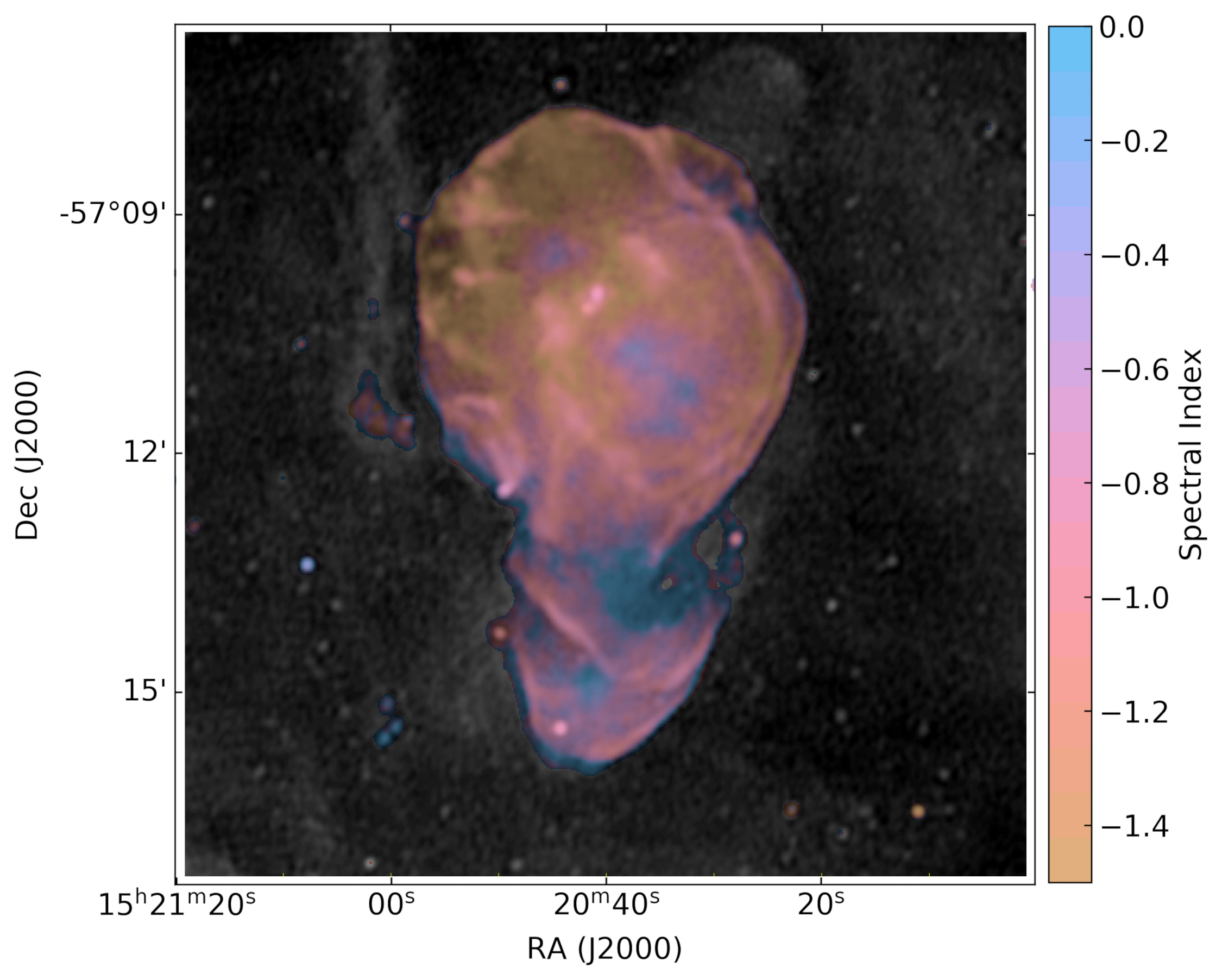}
    \caption{Combined Spectral Index and Intensity. The spectral index was mapped with colours that have the same value (see equilumance mapping at https://github.com/mlarichardson/CosmosCanvas.) This map was blended with a greyscale rendition of intensity using the multiplication mode in GIMP.}
    \label{fig:IntSpecInd}
\end{figure*}

\section{Discussion / Summary}

The jets of Cir X-1 have been identified through radio imaging over the past decades \citep{1993MNRAS.261..593S,1998ApJ...506L.121F,2006MNRAS.372..417T,2011MNRAS.414.3551M,2012MNRAS.419..436C,2019MNRAS.484.1672C}. In \cite{1993MNRAS.261..593S} $\&$ \cite{2006MNRAS.372..417T} the emissivity and curved structures of the jets are detailed, while \cite{2019MNRAS.484.1672C} further revealed additional features in the nebula. Our deep MeerKAT L band image shows these features in much greater detail, such as the rings and bright lines in the southern tip and shell structures. In this work, we make note of new structures in the nebula, such as the diffuse jet-punched bubbles protruding from the position of the rings. These bubbles are positioned in line with the jets of the system. 
To understand the morphology and physical mechanisms responsible in the object, the spectral index maps were determined and the formation of the bubbles simulated. 
\subsection{The spectral index and how it compares to the simulation results}

At the position of the binary system the core is steep with a spectral index around -0.8 in Fig.~\ref{fig:specInd} and light pink-purple in Fig. ~\ref{fig:IntSpecInd}. The core along with the shock fronts of the jet also have the least error according to the spectral index error map (right panel of Fig.~\ref{fig:specInd}). The spectral index appears to steepen as one moves from the core to the rest of the nebula (shell). The region between the core and the shock fronts are optically thin, thus emitting synchrotron radiation, indicating loss of energy and deceleration of the electrons. Once the particles/electrons are confronted by the shock front, some distance away from the core, they experience a surge of energy as they move through the currently observed jets. The shock fronts and the jets therefore appear to have a similar spectral index that is slightly steeper than the core as seen in Fig. ~\ref{fig:IntSpecInd} in pink. These features remain identifiable as flatter than the space around them, which is gold in Fig. ~\ref{fig:IntSpecInd}. Perpendicular to the jets, are much flatter regions with a spectral index more positive than -0.8, blues in Fig. ~\ref{fig:IntSpecInd}. The SW `pocket' is larger and the spectral index is flatter as the shell in its entirety appears to flatten from the north eastern end to the south western end. 
A portion of the NW ring at the edge of the nebula has a spectral index more positive than -0.8 (around -0.1 Fig.~\ref{fig:specInd}), indicating a flatter region compared to the core whereas the SE ring has a spectral index similar to the core. The error in the NW ring spectral index is also quite significant compared to the error of the SE ring. Therefore, the SE ring may be more indicative of the actual spectral index at the rings. 

In the fiducial simulation, once the fast fixed-axis jets turn off and no longer drive the bubbles, it is seen that the bubbles are cooler than the inner nebula. This cooler nature, along with the low surface brightness and density of the bubbles may be the reason there is no spectral index measurement of the bubbles. The simulation evidence of the fast fixed-axis jets being launched early on, during the supernova explosion, and having since turned off (with only the slow less-powerful precessing jets remaining), is consistent with the steep spectral index observed in the radio image over the majority of the nebula (shell). Although the simulations suggest the fast fixed-axis jets being off and there are no longer active jets, regular outbursts could feed into slower precessing jets. This is supported by the spectral index being slightly flatter for the shock fronts and the slower precessing jets compared to the region of space between the core and the jets/shock-fronts. Furthermore, the simulations (both fiducial and others) reproduce the bright rings observed in the radio image and the bright features (rings, core) consistently show a flatter spectral index compared to the shell of the nebula outside the pockets. For the rings, this may be due to the interaction when the jets punched at the edge of the supernova remnant and subsequent ISM interaction. 
\subsection{The minimum energy }

The integrated energy of the jet over its lifetime is $1.1\times10^{50}$erg, much higher than the minimum energy of $4\times10^{45}$erg inferred from the NW bubble in the radio image. In addition, the initial simulated supernova energy is $3\times10^{50}$\,erg, also much higher than the inferred $2\times10^{47}$\,erg minimum energy of the nebula. This is somewhat expected, as both the remnant and bubbles have had time to cool since the energy was injected in the simulation. However, we might expect that the ratio of energies would be the same between the energy injected in the simulation and the minimum energy inferred from the radio image, which currently differ by about a factor of 10.

We choose not to directly compare the integrated flux of the simulated ray-traced image to the 
observed radio image, but instead compare the ratio of the supernova nebula flux to the bubble flux, which we believe to be a more reliable comparison as it is less dependent on scaling freedoms in the flux calculation and assuming that the supernova flux is reliable. We are able to reproduce this contrast, meaning that the discrepancy between the injected and inferred energies comes from either the jet energy being lost at a faster rate than the supernova energy, or the jet emission partially contributing to the nebula flux. In Figure \ref{fig:multisim_comparison}, jets launched early (fiducial and `increase $\Delta t_{\rm jet}$' simulations) slightly over-inflate the supernova nebula in the north and south compared to the late-launched jet (`increase $t_{\rm jet}$' simulation). This could be due to some of the jet energy being dissipated within the nebula itself at early times, thus requiring a larger jet energy to reproduce the nebula-bubble flux ratio. At late times, this energy is still dissipated inside of the nebula but instead inflates a hot inner cocoon as seen in Figure \ref{fig:multisim_comparison}, independent from the outer supernova shock, whereas the jet is able to inflate the nebula itself when launched early. 

Overall, by reproducing the flux contrast between the nebula and bubble, we are showing that adiabatic cooling, in addition to the jet energy redistributing into the inner nebula, is sufficient to explain the discrepancy between the calculated minimum energy and the jet injected energy.

\subsection{Jet properties: considerations and caveats}

The fiducial simulation presented in this work which best reproduces our observations implies the existence of a powerful jet launched early in the evolution of the SN. 
The power of the jet in the simulation is $\sim35L_{\rm edd}$, which is particularly high for a typical NS XRB. However, it has been speculated that SS433, albeit a BH XRB, is a hyper-Eddington ultra-luminous X-ray source which appears comparatively faint due to inclination and collimation \citep{begelman2006nature}. We can therefore postulate that perhaps Cir X-1 is a less evolved sibling of SS433 and is showing analogous super-Eddington behaviour. 
However, we caution that the exact jet power implemented in the simulation is not a `fit' to data, so to speak, and the results simply indicate that a high jet power ($\sim$tens of $L_{\rm edd}$) is a plausible explanation for the radio morphology.

There are also several effects which we have not incorporated in this simulation. Most obviously, although we assume the presence of magnetic fields through synchrotron radiation, we do not invoke them in the simulation. It is possible that large-scale magnetic fields both in the jet itself and the SN nebula could affect the morphology and growth of the bubbles, notably their confinement at the base of the bubble.

In addition, the galactic density profile which we incorporate is a basic model and does not account for local inhomogeneities in the field which, as apparent in our wide-field image (Figure \ref{fig:roughIntensity}), are complex and abundant. These local density fluctuations, in particular a possible underdense region to the North of the nebula and/or an overdense region south to the south, could explain the observed asymmetry in the bubbles. 

Lastly, we have only explored one formation scenario: a `two-mode' jet model where a fast fixed-axis jet is launched in contrast to the present-day slow precessing jet. Although we find that this seems to explain much of our observations, we can not rule out all other jet histories. For example, one possibility is that jet activity before or during the SN explosion could be responsible for the morphology, perhaps from a gamma-ray burst (GRB). However, a GRB jet seems unlikely to explain the observed morphology. We require that the jet remains collimated to large scales in order to produce the bubble height, whereas GRBs are estimated to only remain collimated up until maximally the Sedov length ($\sim 10^{17}$ cm), and in reality are likely to undergo lateral spreading much earlier than this \citep{matthews2025,duffell2018,granot2012}. Indeed, in the simulations of \citet{zhang2009} it is estimated that only after 150 years of evolution the outflow is almost spherical with a radius already about half the size of the final supernova nebula. We note that \citet{shishkin2024identifying} and \citet{2025RAA....25c5008S} have explored the possibility of bubble formation in other SNRs through alternative jittering jet SN mechanisms.

\section{Conclusions}
We produced a deep-field radio image of Cir X-1 using data acquired with the MeerKAT telescope. The image was constructed with a stack of 30 epochs (a total of 7.5\,hrs) such that two full outburst cycles of Cir X-1 were included. The deep radio image revealed a never before seen set of asymmetric jet-punched bubbles. Illustrating a breakout of the jets at the edge of the supernova remnant, similarly seen in the older source, SS433 with supernova remnant W50. We investigate the physical nature of the object from the observational results and simulate the morphological structure to determine the influence of the jets in producing what we observe. We determine that the spectral index map suggests the jets and remnant are largely optically thin. We measure the spectral index of the core presents as slightly flatter than the jets due to the outbursting activity and the `pocket' areas perpendicular to the jets are even flatter (with slightly larger error in these regions compared to the core and shock front/jets combination).

We ran relativistic hydrodynamic simulations of a simplistic supernova explosion, after which we injected a fixed-axis jet which interacts with the expanding SN shell, and explored jet parameters to try to match the observed radio morphology.

Our results suggest that a powerful jet was launched within the first hundred years post-supernova, and that this jet was particularly powerful, on the order of tens of $L_{\rm Edd}$. We can also conclude that the jet was likely short-lived and constant in power, with respect to the timescales of the remnant evolution. We unable to reproduce two main features: the height of the bubble and the narrowness of the base of the bubble i.e. the ring. The bubble height could likely be increased by a locally under-dense region at the north of the nebula, also providing the extra asymmetry that we observe in the radio image. This kind of low-density region could be a natural result of the complex density field found in the galactic plane. 
\\
The narrowness of the ring observed was not reproduced by any test simulation -- this could be due to limitations of the simulation setup itself. Due to the range of scales involved in the system, the width of the jet (although only a few pixels across) is much larger than the width of the binary system and this could potentially contribute to a larger puncture in the nebula at early times, but we did not test this and thus the effects are unknown.  

Overall, we are able to reconstruct a simplified morphology of Cir X-1 with an early-launched powerful fast fixed-axis jet which produced energies as high as 10$^{50}$\,erg. This modeling indicates that the MeerKAT observations are the first to reveal an initial breakout of neutron star
jets from their natal supernova remnant, and further support the scenario in which Cir X-1 is a younger relation of the archetypal
jet source SS433. The detail of the internal structure of the nebula may be investigated in future works. Additionally, the improved telescope facilities like the new MeerKAT S-band (1.75 – 3.5\,GHz) allows for analysis into the jets of the system and their precession (Cowie et al. 2025, in prep) which may help explain the detailed features within the nebula.
\section*{Acknowledgements}
KVSG acknowledges the University of Cape Town, the National Research Foundation Phd scholarship and the Department of Science and Innovation SAWISA Doctoral Fellowship. 
KS acknowledges support from the Clarendon
Scholarship Program at the University of Oxford and the Lester
B. Pearson Studentship at St John’s College, Oxford. JHM acknowledges funding from a Royal Society University Research Fellowship (URF\textbackslash R1\textbackslash221062).

The authors would like to acknowledge Fernando Camilo for helping name the Africa nebula and Wenfei Yu for the valuable discussion in improving the paper.

We gratefully acknowledge the use of the following software packages: \textsc{pluto} \citep{mignone_pluto_2007}, matplotlib \citep{Hunter:2007},  and astropy:\footnote{\url{http://www.astropy.org}} \citep{astropy:2013, astropy:2018, astropy:2022}.

The authors would like to acknowledge the use of the University of Oxford Advanced Research Computing (ARC) facility in carrying out the simulation work (\url{http://dx.doi.org/10.5281/zenodo.22558}).

The MeerKAT telescope is operated by the South African Radio Astronomy Observatory, which is a facility of the National Research Foundation, an agency of the Department of Science and Innovation.
We acknowledge the use of the ilifu cloud computing facility – www.ilifu.ac.za, a partnership between the University of Cape Town, the University of the Western Cape, Stellenbosch University, Sol Plaatje University and the Cape Peninsula University of Technology. The ilifu facility is supported by contributions from the Inter-University Institute for Data Intensive Astronomy (IDIA – a partnership between the University of Cape Town, the University of Pretoria and the University of the Western Cape), the Computational Biology division at UCT and the Data Intensive Research Initiative of South Africa (DIRISA).

This work made use of the CARTA (Cube Analysis and Rendering Tool for Astronomy) software (\cite{2021zndo...3377984C} - DOI 10.5281/zenodo.3377984 – \url{https://cartavis.github.io}).
\section*{Data Availability}

The data availability will be subject to the ThunderKAT LSP data release conditions.



\bibliographystyle{mnras}
\bibliography{reference} 




\appendix




\bsp	
\label{lastpage}
\end{document}